\documentclass[preprintnumbers,amsmath,amssymb,floatfix,11pt,prd,onecolumn,superscriptaddress,nofootinbib]{revtex4-1}

\usepackage{graphicx}
\usepackage{dcolumn}
\usepackage{amsmath,amssymb,mathtools}
\usepackage{epstopdf}
\usepackage{verbatim}
\usepackage{amsmath}

\usepackage[colorlinks=true, citecolor=blue, urlcolor = blue, linkcolor= blue, 
bookmarks=true]{hyperref}
\usepackage{comment}

\newcommand{\be}{\begin{equation}}
\newcommand{\ee}{\end{equation}}
\newcommand{\chushi}[1]{ }

\newcommand{\roundN}[1]{ ( #1 ) }

\newcommand{\braN}[1]{ \langle #1 | }
\newcommand{\ketN}[1]{ | #1 \rangle }

\let\calccommentout\iffalse 
\let\calcshow\iftrue 

\labelformat{section}{Section #1} 
\labelformat{subsection}{Section #1} 
\labelformat{subsubsection}{Section #1}
\labelformat{subsubsubsection}{Section #1}
\labelformat{equation}{Eq.~(#1)} 
\labelformat{figure}{Fig.~#1} 
\labelformat{subfigure}{Fig.~\thefigure#1} 
\labelformat{table}{Tab.~#1} 
\labelformat{appendix}{Appendix #1}

\begin{document}

\title{Entanglement generation between  
Unruh-DeWitt detectors in the de Sitter spacetime -- analysis with complex scalar fields }
\author{Shagun Kaushal}
\email{shagun.kaushal@vit.ac.in, shagun123@iitd.ac.in}
\affiliation{Department of Physics, Vellore Institute of Technology, Vellore, Tamil Nadu 632 014, India\footnote{Current affiliation}}
\affiliation{Department of Physics, Indian Institute of Technology Delhi, Hauz Khas, New Delhi 110 016, India}

\author{Sourav Bhattacharya}
\email{sbhatta.physics@jadavpuruniversity.in}
\affiliation{Relativity and Cosmology Research Centre, Department of Physics, Jadavpur University, Kolkata 700 032, India}


\begin{abstract}
\noindent
We investigate the entanglement generation or harvesting between two identical, comoving Unruh-DeWitt detectors in the cosmological de Sitter spacetime. The detectors are assumed to be unentangled initially. They are individually coupled to a complex scalar field, which eventually leads to coupling between themselves. Two kinds of complex scalar fields are investigated here — conformally invariant and massless minimally coupled. By tracing out the degrees of freedom corresponding to the scalar, we construct the reduced density matrix for the two detectors, whose eigenvalues characterise transition probabilities between the energy levels of the detectors. We have computed the negativity, quantifying the degree of entanglement generated at late times between the two detectors. The similarities and differences of these results between the aforementioned two kinds of scalar fields have been discussed.  We also compare our results with the existing result of the real scalar field, and point out the qualitative differences.   In particular, we emphasise that entanglement harvesting is more resilient in scenarios involving complex fields and nonlinear couplings.  
\end{abstract}
\maketitle

\newpage

\tableofcontents 

\section{Introduction}\label{s1}

\noindent
The phenomenon of quantum entanglement, owing to its  non-local characteristics, is considered to be  even more counter intuitive compared to the standard quantum mechanical processes. Experimental observations, for example, of the violation of the Bell inequalities, which cannot be explained by classical theories based on  hidden variables, have placed quantum entanglement on very strong physical grounds~\cite{bell_1, bell_2, bell_3, bell_4}. One of the defining characteristics of entangled states is the inability to describe the corresponding Hilbert space as a product of pure states of subsystems, e.g.~\cite{NielsenChuang, Horodecki:1997vt} and references therein.  In addition to its foundational role in quantum physics, quantum entanglement has already found a wide range of applications in e.g., building high security communication systems by employing quantum cryptography to futuristic quantum teleportation-based devices, e.g.~\cite{WJHN, DAC, Hotta:2008uk, MKM, Fuentes-Schuller:2004iaz}.

Recently, the research community has  shown considerable interest in studying quantum entanglement in the context of relativistic quantum field theory~\cite{Reznik:2002fz, Martin-Martinez:2015qwa, Fuentes:2010dt, Anastopoulos:2022owu, Casini:2022rlv, Preskill, RevModPhys, Jordan:2011ci, Giddings:2012bm, Calabrese:2004eu, Wu:2022glj, K:2023oon}. An essential focus of this topic  is to examine the dynamics of particles coupled to a quantum field, particularly using particle detectors. These investigations encompass studying entanglement dynamics, entanglement harvesting, and understanding the radiative processes of entangled relativistic particles \cite{Menezes:2015veo, Menezes:2015iva, Iso:2016lua, Lindel:2023rfi, Lima:2023pyt, Elghaayda:2023igv, Barman:2022xht, Bhattacharya:2020sjr, Kaushal:2024cbk, Kaushal:2022las}. Entanglement harvesting in particular, can be highly exciting, for it can offer  means to extract additional quantum information.

 The Unruh-DeWitt detectors are very popular in the context of relativistic quantum entanglement. Initially designed for studying radiation observed by a uniformly accelerated observer in the Minkowski spacetime, they were also used for studying Hawking radiation in eternal black hole spacetimes~\cite{Unruh} (see also~\cite{Birrell:1982ix} and references therein). In this paper, we wish to examine the conditions under which these detectors become entangled over time in the presence of a coupling with a quantum field; such a framework allows us to explore the potential for entanglement harvesting between initially uncorrelated detectors. Various earlier investigations show different factors could contribute to this detector entanglement, including detectors' trajectories~\cite{Reznik:2002fz, Benatti:2004ee, Salton:2014jaa, Koga:2018the, Koga:2019fqh, Zhang:2020xvo}, the background spacetime geometry~\cite{Fuentes-Schuller:2004iaz, Henderson:2017yuv, Wu:2023glc, Martin-Martinez:2012chf, Tjoa:2022oxv, Bhattacharya:2022ahn, Dhanuka:2022ggi, Maeso-Garcia:2022uzf, Martin-Martinez:2015eoa, Mendez-Avalos:2022obb, Perche:2022ykt}, the presence of a thermal bath~\cite{Brown:2013kia, Barman:2021kwg, Henderson:2022oyd}, and even the transient passage of gravitational waves~\cite{Prokopec:2022akz, Xu:2020pbj, Barman:2023aqk, Nambu:2013rta, Kukita:2017etu, Kukita:2017etu}. 

A large number of works have actively engaged in understanding the entanglement harvesting patterns of Unruh-DeWitt detectors that interact perturbatively with quantum fields. These works span from inertial detectors in flat spacetime~\cite{Barman:2022xht, Barman:2021kwg, Koga:2018the, Koga:2019fqh} to those following various trajectories in curved spacetimes~\cite{Fuentes-Schuller:2004iaz, Martin-Martinez:2015qwa, Fuentes:2010dt, Henderson:2017yuv, Robbins:2020jca, Cong:2020nec, Gallock-Yoshimura:2021yok}. For various such studies, we further refer our reader to~\cite{Reznik:2002fz, Benatti:2004ee, Salton:2014jaa, Zhang:2020xvo, Brown:2013kia, Henderson:2022oyd, Pozas-Kerstjens:2015gta, Xu:2020pbj, Cliche:2010fi} (also references therein). See~\cite{Chowdhury:2021ieg, Barman:2022loh} for study of  degradation/entanglement generation between two initially correlated Unruh-DeWitt detectors. See also~\cite{Makarov:2017erw, Brown:2012pw} for discussion on some non-perturbative effects. We note that studying such features of entanglement in the early inflationary background can provide valuable insights about the geometry as well as the quantum condition at such stage.  For instance, entanglement generated in the early universe could affect cosmological correlation functions or the cosmic microwave background~\cite{Boyanovsky:2018soy, Rauch:2018rvx, Morse:2020mdc}.  Researchers are increasingly interested in investigating aspects of entanglement in the cosmological de Sitter spacetime, for more details, readers may refer to \cite{Bhattacharya:2019zno, Bhattacharya:2020bal, Ali:2021jch, Foo:2020dzt, VerSteeg:2007xs, Brahma:2023hki, Brahma:2023lqm, Bhattacharyya:2024duw, Bousso:2001mw, Garbrecht:2004du, SKS}.

In this work, we focus on examining entanglement harvesting between two two-level, identical, initially unentangled  Unruh-DeWitt detectors coupled to a complex scalar field in the cosmological de Sitter spacetime. Note that for the complex scalar, the hermiticity of the field-detector coupling makes it qualitatively different from the real scalar. We take the trajectories of these detectors to be comoving, i.e., their spatial positions are fixed. However, their proper separation will increase with time due to the spacetime expansion. We investigate two physically interesting scenarios: a conformal complex scalar in the conformal vacuum and a minimally coupled complex massless scalar. We assume that the detectors are initially at ground state, whereas the initial state of the field is the vacuum. By constructing the appropriate reduced density matrix for the detectors at the leading perturbative order, we next compute the logarithmic negativity for each scenario to quantify the entanglement generated or harvested between the detectors. Furthermore, we explore how the characteristics of this harvested entanglement vary with different system parameters, including their spatial separation. Single Unruh-DeWitt detector response functions for the above kinds of scalars in the de Sitter background can be seen in~\cite{Bousso:2001mw, SKS}.

The rest of this paper is organised as follows. In the next section, we briefly review the basic framework. \ref{int} focuses on the entanglement generation or harvesting for detectors coupled to complex conformal and massless minimal scalar fields, by computing the negativity. 
Finally, we conclude in \ref{Conclusion}. \ref{A} and~\ref{A1} provide additional details of our calculations, following and in some instance extending the method outlined in~\cite{Nambu:2013rta} for a real scalar field theory. We will work with the mostly positive signature of the metric and will set $c=1=\hbar$ throughout. 

\section{The basic setup} \label{setup}

\noindent
Following~\cite{Unruh, Birrell:1982ix, Bousso:2001mw, Garbrecht:2004du, SKS, Allen:1985ux}, we wish to sketch below the basic setup we will be working in, for the sake of completeness. The de Sitter metric in expanding cosmological coordinates reads
\begin{eqnarray}
ds^2= - d\tau^2 + e^{2H \tau} \left(dx^2+dy^2+dz^2\right)= \frac{1}{H^2 \eta^2}\left[- d\eta^2 + dx^2+dy^2+dz^2\right]
\label{d1}
\end{eqnarray}
where $H>0$ is the de Sitter Hubble constant, and  $\eta = - e^{-H\tau}/H$, is the conformal time.

The general action for a  
complex scalar field reads
\be
S= -\frac12\int \sqrt{-g}\, d^4 x \,\left[ (\nabla_{\mu}\phi^{\dagger})(\nabla^{\mu}\phi) +(m^2 + \xi R) |\phi|^2 \right]
\ee
We will be concerned with two scenarios in this paper, viz., a conformally invariant complex scalar field ($m^2=0,\ \xi  =1/6$) and a massless minimally coupled complex scalar ($m^2=0=\xi $).

\noindent
Let us now quickly review the Unruh-DeWitt particle detector formalism, e.g.~\cite{Unruh, Birrell:1982ix}.  For a real scalar field, the simplest coupling with a point-like detector reads
\be
{\cal L}_{\rm int, real}= - g \mu (\tau) \phi (x(\tau))
\label{R}
\ee
where $\mu$ is detector's monopole moment operator, $g$ is the field-detector coupling constant, and $\tau$ is the cosmological time along the trajectory of the detector. In the Heisenberg picture, we have $\mu(\tau)= e^{iH_0 \tau }\,\mu\, e^{-iH_0 \tau}$, where $H_0$ is detector's free Hamiltonian. In this paper, we wish to restrict ourselves to comoving trajectories, i.e., we assume that the spatial position of the detector is fixed. However, their physical separation will change due to spacetime expansion.

For a transition from an initial state to a final state $| i\rangle \to | f \rangle$ in this  system, the first-order transition matrix element reads
\begin{eqnarray}
{\cal M}_{fi} = ig \,\langle \omega_f | \mu | \omega_i \rangle \int_{\tau_i}^{\tau_f} d\tau e^{- i (\omega_f-\omega_i) \tau } \langle \phi_f | \phi(x(\tau)) | \phi_i\rangle
\label{d4}
\end{eqnarray}
where we have taken $| i\rangle = | \omega_i \rangle \otimes| \phi_i \rangle$ and $| f\rangle = | \omega_f \rangle \otimes | \phi_f \rangle$, where  $\omega$'s are the energy eigenvalues of the detector. The transition probability is given by 
\be
|{\cal M}_{fi}|^2=g^2 \,|\langle \omega_f | \mu | \omega_i \rangle|^2 \int_{\tau_i}^{\tau_f} d\tau_1\, d\tau_2 \,e^{- i (\omega_f-\omega_i) (\tau_1-\tau_2) }\, \langle \phi_i | \phi(x_2(\tau_2)) | \phi_f\rangle\langle \phi_f | \phi(x_1(\tau_1)) | \phi_i\rangle 
\ee 
We next sum over all the  final states of the field,  using the completeness relation for the field basis, $|\phi_f \rangle$. Assuming the initial state for the field to be the vacuum, we have the transition probability 
\begin{eqnarray}
 {\overline {\cal F}(\Delta \omega)}= \int {\cal D} \phi_f|{\cal M}_{fi}|^2=g^2 \,|\langle \omega_f | \mu | \omega_i \rangle|^2 \int_{\tau_i}^{\tau_f} d\tau_1\, d\tau_2 \,e^{- i (\omega_f-\omega_i) (\tau_1-\tau_2) }\, \langle \phi_i | \phi(x_2(\tau_2))\phi(x_1(\tau_1)) | \phi_i\rangle \nonumber\\
 \label{rf}
\end{eqnarray}
where
$$\langle \phi_i | \phi(x_2(\tau_2))\,  \phi(x_1(\tau_1)) | \phi_i\rangle \equiv  iG^+(x_2(\tau_2)-x_1(\tau_1))$$
is the Wightman function. One then defines two new temporal variables
\begin{equation}
    \label{xy}
\Delta \tau =\tau_1-\tau_2\qquad \text{and}\qquad \tau_+ = \frac{\tau_1+\tau_2}{2}
\end{equation}
with ranges $-\infty \ensuremath{<} \Delta \tau \ensuremath{<}\infty $, $-\infty \ensuremath{<}\tau_+ \ensuremath{<}\infty $. However, since the Wightman function usually is not a function of $\tau_+$, \ref{rf} is divergent. In order to thus give it a physical meaning, one defines transition probability per unit proper time by dividing/differentiating it by $\tau_+$~\cite{Unruh}. The resulting transition probability per unit proper time is known as the  response function, given by 
\begin{eqnarray}
 \frac{d {\cal F}(\Delta \omega)}{ d \tau_+} =  \int_{-\infty }^{\infty} d \Delta \tau \, e^{- 2i  (\omega_f-\omega_i) \Delta \tau} \, i G^{+}(\Delta \tau)
\label{d5}
\end{eqnarray}
where we have abbreviated  ${\cal F} := \overline{\cal F}/(g |\langle \omega_f | \mu | \omega_i \rangle|)^2$.  \\

\noindent
 For a scalar field moving in \ref{d1} of mass $m$ and non-minimal coupling $\xi$, the Wightman function $i G^+(x,x')$ reads~\cite{Allen:1985ux},
\begin{eqnarray}
iG^+(x,x') = \frac{H^2}{16\pi^2}\Gamma \left(\frac32 -\nu\right) \Gamma\left(\frac32 +\nu\right)\, _2F_1 \left(\frac32 -\nu, \frac32 +\nu, 2; 1-\frac{y(x,x')}{4} \right)
\label{d7}
\end{eqnarray}
where 
$$\nu = \left(\frac94 -12 \xi -\frac{m^2}{H^2}\right)^{1/2}$$
and the de Sitter invariant interval $y(x,x')$  reads
$$y(x,x')=\frac{-(\eta- \eta'-i\epsilon)^2+ |\vec{x}-\vec{x}'|^2}{\eta \eta'}$$
where $\epsilon=0^+$. Rewriting things now in the cosmological time $t$ and setting $\vec{x} =\vec{x}'$, appropriate for a comoving  detector, we have
\begin{eqnarray}
y(\tau,\tau')=-4\left(\sinh\frac{H\Delta \tau}{2}-i\epsilon\right)^2
\label{d7''}
\end{eqnarray}
Note that since we have set the comoving spatial separation to zero, the cosmological time $\tau$ becomes the proper time along the detector's trajectory. We note that if more than two detectors are present, we cannot  ignore the spatial coordinates from our calculations when two point function connecting these two points appear, due to their non-vanishing {\it mutual} spatial separation. Hence we will keep it below for our current purpose.\\

\noindent
For a complex scalar field, the simplest  field-detector coupling reads,
\begin{equation}
   {\cal L}_{\rm int}=- g(\tau) \mu (\tau) \phi^{\dagger}(x(\tau)) \phi (x(\tau)) \label{complex} 
\end{equation}
Alike the fermions, e.g.~\cite{Louko:2016ptn}, the quadratic coupling is necessary in order to make the interaction Hamiltonian hermitian. This means that the corresponding response function integrals will contain a product of two Wightman functions. Before moving forward, let us write the Wightman function of a conformally invariant complex scalar field. It can be obtained by setting \( m=0 \) and \( \xi=1/6 \) in \ref{d7}, and is given by
\begin{eqnarray}
    \label{G+conformal'}
    iG_C^+(x,x')=\frac{H^2}{4 \pi^2} \left[ 4\sinh^2{\left(\frac{H}{2}(\Delta \tau-i\epsilon)\right)}+e^{H(\tau_1+\tau_2)}r^2 \right]^{-1},
\end{eqnarray}
where $r =|\vec{x}_1-\vec{x}_2|$ is the comoving separation.
For a massless and minimally coupled scalar field, we have $\nu=3/2$ and hence \ref{d7} has no sensible limit. Thus the Wightman function in this case must be determined individually. 
For our present purpose, 
we will take the form derived in~\cite{Nambu:2013rta}, 
\begin{equation}
    \label{WMCS}
    \begin{split}
   i G_{M}^+(x,x')&=\frac{H^2}{16\pi^2} \left[ -\sinh^2{\left(\frac{H}{2}(\Delta \tau-i\epsilon)\right)}+e^{H (\tau_1+\tau_2)}\left(\frac{Hr}{2}\right)^2 \right]^{-1}  \\ 
    &\quad -\frac{H^2}{8\pi^2} {\rm Ei}\Bigg[-\frac{i\Tilde{\epsilon}}{H} \Bigg(-Hr+2e^{-\frac{H(\tau_1+\tau_2)}{2}}\sinh{\left(\frac{H}{2}(\Delta \tau-i\epsilon)\right)}\Bigg)\Bigg] \\ 
    &\quad -\frac{H^2}{8\pi^2} {\rm Ei}\Bigg[-\frac{i\Tilde{\epsilon}}{H} \Bigg(Hr+2e^{-\frac{H(\tau_1+\tau_2)}{2}}\sinh{\left(\frac{H}{2}(\Delta \tau-i\epsilon)\right)}\Bigg) \Bigg] +\frac{H^2}{4\pi^2}
    \end{split}
\end{equation}
where Ei is the exponential integral function  given by~\cite{GR},
\begin{equation}
    {\rm Ei}[z]=\int_{z}^{\infty} \frac{dt}{t} e^{-t}
\end{equation}
In addition to the usual regulator $\epsilon=0^+$, there is another positive infinitesimal parameter appearing above,  $\Tilde{\epsilon} \ll H$, in order to regularise the infrared divergences. Physically,  ${\Tilde{\epsilon}}^{-1}$ represents the horizon scale at the onset of inflation. We refer our reader to~\cite{Kukita:2017etu} for detailed discussions.

\subsection{Complex scalar field coupled to two Unruh-DeWitt detectors}\label{UD1}

\noindent
We now consider two Unruh-DeWitt detectors coupled to a complex scalar field. As of \ref{R}, the simplest interaction Hamiltonian reads for this composite system as 
\begin{equation}
    \label{HamiltonianR}
    H_{I}=\sum_j g_{j}\mu_{j}(\tau_j) \phi^\dagger(x(\tau_{j})) \phi(x(\tau_{j})) 
\end{equation}
where the index $j$ runs for both the detectors, labeled as $A$ and $B$. We explicitly expand the monopole moment of the detectors (introduced below \ref{R}) in the Heisenberg picture as a function of the proper time as \cite{monopole, Bruschi:2012rx, Sriramkumar:1994pb}
\begin{equation}
        \label{momentR}
        \mu_{j}(\tau_j) =|E_{j}\rangle \langle G_{j} | e^{i \omega_{j} \tau_{j}} + |G_{j}\rangle \langle E_{j} | e^{-i \omega_{j} \tau_{j}}  \qquad ({\rm no~sum})
    \end{equation}
$G$ and $E$ in the above expression stand respectively for the ground and excited levels of the detectors. For simplicity, we imagine both detectors to be identical so that $\omega_j=\omega$ and $g_j=g$. We also assume that both the detectors and the scalar field are in their ground states initially, so that the initial state is given by  
    \begin{equation}
        \label{instateR}
        |\text{in}\rangle=|0 \otimes G_A \otimes  G_B\rangle
    \end{equation}
    Being a perfectly product state, it does not represent any entanglement initially.
From now on, we shall suppress the tensor product symbol for the sake of notational brevity. The time evolution of this initial state in the interaction picture is given by
    \begin{equation}
        \label{outdefR}     |\text{out}\rangle=U|\text{in}\rangle=T e^{-i\int d\tau_j H_{I}(t(\tau_j))}|0 G_A G_B\rangle
    \end{equation}
where $T$ stands for time ordering. We make the perturbative expansion, $U= I+U^{(1)}+U^{(2)}+\cdots$, with 
\begin{eqnarray}
    \label{Uexpn}
    U^{(1)}&=&-i \int_{-\infty}^{\infty}d\tau_j\; H_{I}(t(\tau_j))\\
     U^{(2)}&=&- \int_{-\infty}^{\infty}d\tau_i\int_{-\infty}^{\tau}d\tau'_j\; H_{I}(t(\tau_i))  H_{I}(t^\prime(\tau_j'))
\end{eqnarray}
and so on. The density operator corresponding to the out state is given by  
    \begin{equation}
        \label{densityoutR}
        \rho=|\text{out}\rangle \langle\text{out}|=\rho^{(0)}+\rho^{(1)}+\rho^{(2)}+O(g^3)
    \end{equation}
where $\rho^{(n)}$ is of the order of $g^{n}$. Also note that, since the initial state of the field is the vacuum, we have $\rho^{(1)}=0$ due to the vanishing one point function of the field. The two detectors interact with each other via the scalar field. The reduced density matrix of the two detectors is obtained by tracing out the field $\phi$'s degrees of freedom, resulting in the mixed density matrix
    \begin{equation}
        \label{rhoABR}
       \rho_{\text{\small AB}}
={\rm Tr}_{\phi} \rho= \left(\begin{array}{cccc}
            1-P_{AA}-P_{BB} &0&0&E^*  \\
            0&P_{AA}&P_{AB}&0 \\
        0&P_{AB}&P_{BB}&0 \\
        E&0&0&0
        \end{array}\right )
    \end{equation}
The basis of $\rho_{\text{\small AB}}
$ is $|G_AG_B\rangle,\;|G_AE_B\rangle,\;|E_AG_B\rangle\;\text{and}\;|E_AE_B\rangle$, and the matrix elements explicitly read,
\begin{equation}
    \label{PA}
    P_{IJ}=g^2\int d \tau_I \int d \tau'_J\; e^{-i\omega(\tau_I-\tau'_J)}  (iG^+(x_I,x_J'))^2 \qquad (I,J=A,B)
\end{equation}
\begin{equation}
    \label{E}
  E = - g^2\int d\tau_A \int d\tau_B\;  e^{i\omega (\tau_A+\tau_B)} (iG^+(x_A,x_B))^2
\end{equation}
where $iG^+$ is the positive frequency Wightmann function. 

    Now we compute the negativity to quantify the entanglement between the two detectors represented by \ref{rhoABR}. The negativity of a bipartite state is defined as the absolute sum of the negative eigenvalues of $\rho_{AB}^{\text{T}_A}$,  where $\rho_{AB}^{\text{T}_A}$ is the partial transpose of $\rho_{AB}$ with respect to the subspace of $A$, i.e., $\roundN{ \ketN{i}_{\! A} \hspace{-0.2ex} \braN{n} \otimes \ketN{j}_{\! B} \hspace{-0.2ex} \braN{\ell} }^{\text{T}_A}: = \ketN{n}_{\! A} \hspace{-0.2ex} \braN{i} \otimes \ketN{j}_{\! B} \hspace{-0.2ex} \braN{\ell}$.
The partial transposed density matrix $\rho_{AB}$ reads
\begin{equation}
        \label{rhoABTR'}
      (\rho_{AB})^{{\rm T}_A}=\left(\begin{array}{cccc}
            1-2P &0&0&P_{AB} \\
            0&P&E&0 \\
        0&E^*&P&0 \\
        P_{AB}&0&0&0
        \end{array}\right )
    \end{equation}
where we have used $P_{AA}= P_{BB}=P$, owing to the identical nature of the detectors. The eigenvalues of \ref{rhoABTR'} are given by
\begin{eqnarray}
    \label{EV1}
    \lambda_1&=&\frac{1}{2}\left( 1-2P+\sqrt{(1-2P)^2+4P^2} \right) \nonumber\\
    \lambda_2&=&\frac{1}{2}\left( 1-2P-\sqrt{(1-2P)^2+4P^2} \right) \nonumber\\
    \lambda_3&=&P+|E|\qquad    \lambda_4= P-|E|
\end{eqnarray}
Now we recall the fundamental result that a bipartite system which satisfies 
\begin{equation}
    \label{cond}
    |E|\geq P,
\end{equation}
is entangled \cite{Horodecki:1997vt}. In this case, the negativity is defined as~\cite{Vidal:2002zz, Plenio:2005, Calabrese:2012nk},
\begin{eqnarray}
    N= \text{max}[0,|E|-P]
\end{eqnarray}
Therefore, for our present case we have  
\begin{equation}
    \label{neg}
    N=|E|-P
\end{equation}
For initially separable detector states, a positive value of \(N\) following interaction with the scalar field suggests that the detectors become entangled finally. 
Negativity serves as a necessary and sufficient criterion for entanglement in two-qubit systems~\cite{Horodecki:1997vt}.\\

\noindent
The transition integrals of \ref{PA}, \ref{E} are not convergent, owing to the integral over the variable $\tau_+$, \ref{xy}. In order to regularise these integrals, one introduces a switching or window function, e.g.~\cite{Nambu:2013rta}, and references therein. Such function effectively restricts the interaction time between the detectors. The most popular, which we will also  be using, is the Gaussian window function,
\begin{equation}
    \label{GTAU}
    g \to g(\tau):=g e^{-\tau^2/2\sigma^2}
\end{equation}
in \ref{PA} and \ref{E}.
This effective interaction profile ensures a smooth interaction transition.

In de Sitter spacetime, two comoving detectors are causally disconnected if their comoving distance \(r\) satisfies
\begin{equation}
    r \ensuremath{>} \frac{2\sinh{H\sigma}}{H},
\end{equation}
 This condition arises because the comoving causal length  for each detector is given by
\begin{equation}
    r_{\text{null}}=\int_{-\sigma}^{\sigma}\frac{d\tau}{a} = \frac{2\sinh{H\sigma}}{H}.
\end{equation}
Thus for two detectors initially in a separable state and satisfying the condition, \begin{equation}
    \label{Ieq}
    r > r_{\text{null}},
\end{equation}
the entanglement harvested can be attributed to the intrinsic entanglement of the scalar field, as local quantum measurements cannot generate entanglement between causally disconnected regions. Conversely, if $r < r_{\text{null}}$, the detectors are causally connected, allowing for direct causal interaction. In this case, attributing the observed entanglement as solely arising from the non-local correlations of the scalar field becomes less straightforward due to the possibility of the local interactions.\\

\noindent
As we have stated above, if we do not use a switching function, the interaction persists indefinitely, leading to divergences in the transition probability due to the infinite integration over time. This also results in the appearance of the  logarithmic secular growth for a massless minimally coupled scalar field, e.g.~\cite{Miao:2010vs, SKS} and references therein. To address these divergences and extract physically meaningful quantities, one alternative approach would be to compute transition probabilities per unit proper time. This is done by redefining the temporal variables as of \ref{xy}
and then normalising the transition  integrals by dividing it by \(\lim_{\tau_+ \to \infty} \tau_+\). This prescription ensures a well-defined transition rate, preventing unphysical divergences. However while it does so, it is not difficult to verify that the absence of a switching function still leads to ill-defined energy eigenstates and density matrix elements. Therefore, introducing a switching function seems to be essential to overcome these effects and ensure well defined transition rates and density matrix eigenvalues. Nevertheless, a plausible critique to the introduction of such function still will be, in a cosmological scenario, turning on or off the interaction can never be in our hand. In fact it might be possible that the interaction remains omnipresent, as is always the case in a non-equilibrium scenario.  

\noindent
With these, are now ready to go into the computation of entanglement generation due to field-detectors couplings.

\section{Computation of the transition integrals and the negativity}\label{int}
\subsection{A complex conformal scalar field }\label{qcs}
\noindent
Let us begin with a conformally invariant complex scalar field in the conformal vacuum, coupled to two Unruh-DeWitt detectors. Substituting \ref{G+conformal'}, \ref{GTAU} into \ref{PA} and \ref{E}, we have the following transition integrals
\begin{eqnarray}
\label{PEC}
&&P_{IJ} =  g^2\int d \tau_I d \tau'_J \ e^{-(\tau_I^2+{\tau'}_J^2)/2\sigma^2} e^{-i\omega(\tau_I-{\tau'}_J)} (i G^+_C(x_I,x'_J))^2 \qquad (I,J=A,B)\nonumber\\ &&
E= -g^2 \int d \tau_A  d \tau'_B \ e^{-(\tau_A^2+{\tau'}_B^2)/2\sigma^2} e^{-i\omega(\tau_A+{\tau'}_B)} (i G^+_C(x_A,x'_B))^2,
\end{eqnarray}
By substituting these expressions into \ref{neg}, we can determine the negativity for the two identical detectors. In order to compute the negativity, it is necessary to evaluate \( P_{IJ} \) for \( I = J \) (i.e. the same detector), where the spatial separation  is set to zero \( r = 0 \). We have for identical detectors
\begin{equation}
\begin{split}
    \label{Pconf}
   P_{AA}=P_{BB}= P(r=0) 
   =\frac{g^2 H^4}{256 \pi^4} \int d \tau \int d \tau' \frac{e^{-\frac{\tau^2+\tau'^2}{2\sigma^2}}e^{-i\omega(\tau-\tau')}}{\big[\sinh{\left(H\left(\frac{\tau-\tau'}{2}-i\epsilon\right)\right)}\big]^4}
    \end{split}
\end{equation}
On rewriting this integral in terms of temporal variables defined in \ref{xy}, we have
\begin{equation}
    \label{Pxy}
    P= \frac{H^4 g^2}{256 \pi^2}\int_{-\infty}^{\infty} d\tau_+ \;e^{-\tau_+^2/\sigma^2}\int_{-\infty}^{\infty} d (\Delta \tau) \frac{e^{-\Delta \tau^2/\sigma^2}e^{-2i\omega \Delta \tau}}{\big[\sinh({H(\Delta \tau-i\epsilon)})\big]^4},
\end{equation}
which, after  rescaling the temporal variables, and 
defining $H\sigma=h$,    is rewritten as
\begin{equation}
\label{Pxy2}
\begin{split}
P&=\frac{H^2 g^2 h^2}{256 \pi^4} \int_{-\infty}^{\infty}dx\; e^{-x^2}\int_{-\infty}^{\infty}dy \frac{e^{-y^2-2i\omega \sigma y}}{\big[\sinh({h(y-i\epsilon)})\big]^4}=\frac{H^2 g^2 h^2}{256 \pi^{7/2}} \int_{-\infty}^{\infty}dy \frac{e^{-y^2-2i\omega \sigma y}}{\big[\sinh({h(y-i\epsilon)})\big]^4}
    \end{split}
\end{equation}
Likewise, the second of \ref{PEC}  gives,
\begin{equation}
    \label{Exy}
    E= -\frac{H^2 g^2 h^2}{128 \pi^4}\int_{-\infty}^{\infty}dx\; e^{-x^2+2i\omega\sigma x}\int_{-\infty}^{\infty}dy \frac{e^{-y^2}}{\big[\sinh^2{(h(y-i\epsilon))}-e^{2hx}(Hr/2)^2\big]^2}
\end{equation}
We could not evaluate \ref{Pxy2} and \ref{Exy}  in closed forms and have found them numerically. However, before we do that, we need to cast them into a more tractable form, as described in \ref{A}. The reformulated integrals are respectively given by
\begin{equation}
    \label{PCfinal}
  P  = \frac{H^2 g^2 h^2e^{-\omega^2 \sigma^2}}{96\pi^3} \int_{0}^{\infty} dk \; e^{-k^2} k^3 \frac{\cosh{\frac{k\pi}{h}(1-\frac{2h\omega \sigma}{\pi})}}{\sinh{\frac{k\pi}{h}}}
\end{equation}
\begin{equation}
    \label{ECfinal}
E=-\frac{H^2 g^2}{64 \pi^{3/2}}\int_{-\infty}^{\infty}dx\; e^{-x^2+2i\omega \sigma x} \int_{-\infty}^{\infty} dk\; e^{-k^2}  \frac{ \sin{\Big(\frac{2k\ln{b}}{h}\Big)} \cosh{\frac{2\pi k}{h}}}{a\left(1+a^2h^2\right)\left(1-e^{\frac{2\pi k}{h}}\right)}\Bigg(\frac{h}{\ln{b}}+2ik\Bigg) 
\end{equation}
where we have abbreviated, $a=e^{hx}r/2\sigma$ and $b=ah+\sqrt{a^2 h^2 +1}$.
Following~\cite{Nambu:2013rta}, we next assume the following parameter ranges for asymptotic numerical analysis: 
\begin{equation}
    \label{ranges}
    \omega \sigma \gg 1,\; \frac{\pi}{h} \gg 1, \;\text{and}\; \frac{r}{2\sigma} \gg  1
\end{equation}
The first simply states that the characteristic time scale associated with the detectors are much small compared to their interaction time. Since $h=H\sigma$, the second condition implies that the interaction time scale is much small compared to that of the  inflation. The third is an outcome of the fact that $h \ll 1$, and it indicates that the two detectors are causally disconnected. For a detailed calculation of the asymptotic analysis, we refer the reader  to \ref{A1}.

\begin{figure}
    \centering
    \includegraphics[width=1.0\linewidth]{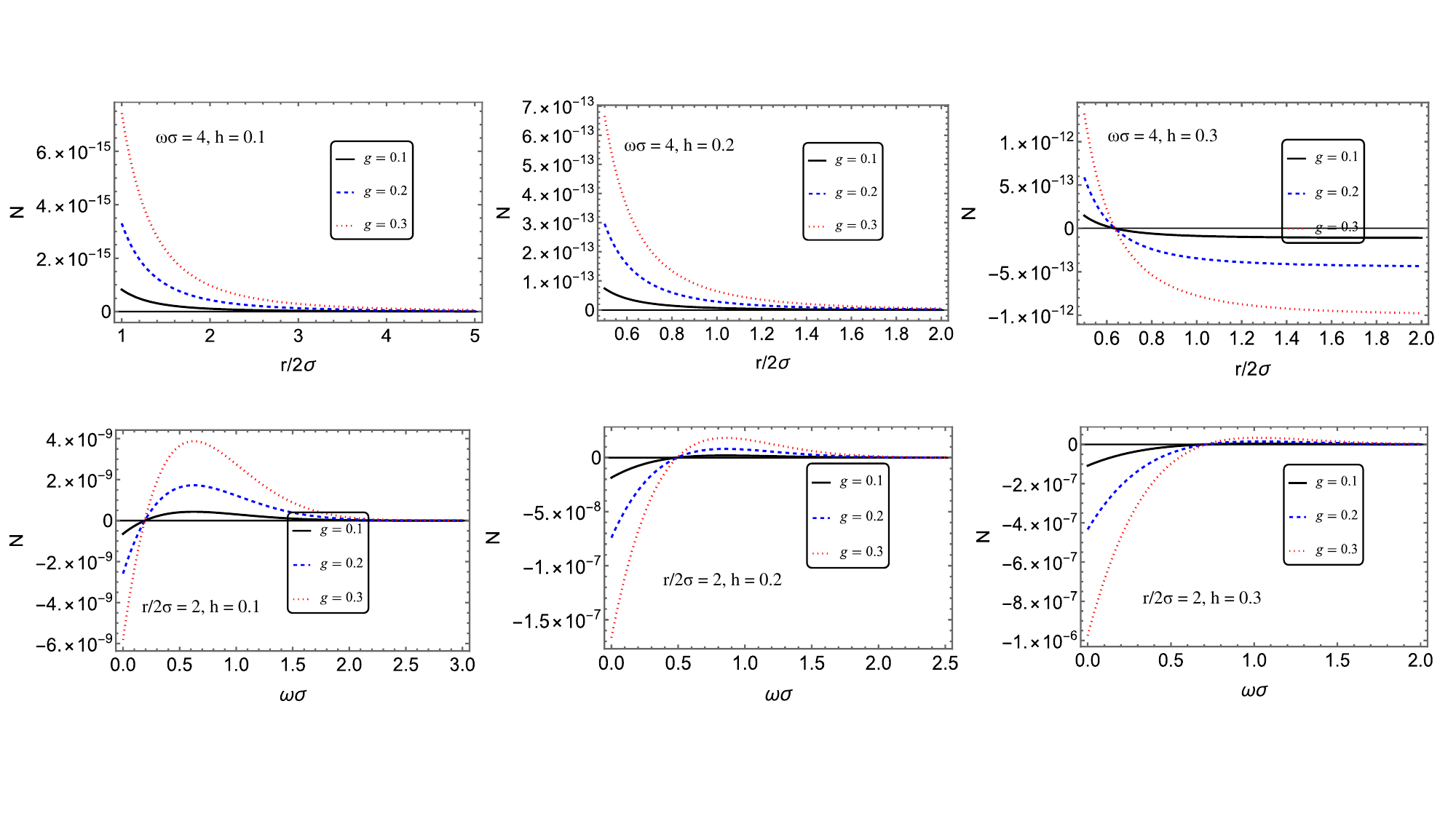}
    \vspace{-1.5cm}
    \caption{\small \it Negativity between two Unruh-DeWitt detectors coupled to a conformally invariant, massless complex scalar field as a function of the dimensionless parameters $r/2\sigma$ and $\omega \sigma$, for various values of coupling $g$ and interaction duration $h = H\sigma$. Note that a negative $N$ implies that the physical negativity is zero, and hence no entanglement  between the detectors.}
    \label{N1}
\end{figure}

After evaluating \ref{PCfinal} and \ref{ECfinal} numerically, we have computed the entanglement negativity using  \ref{neg}.\ref{N1} shows the behaviour of the same, with respect to  the dimensionless parameters $r/2\sigma$ (detectors' separation) and $\omega \sigma$ (energy gap), for various values of the coupling strength $g$ and interaction duration $h = H\sigma$. As can be seen, the negativity can be nonzero even for super-horizon separations ($r/2\sigma \ensuremath{>} 1$), revealing persistent non-local correlations in the conformal vacuum. Increasing $h$ suppresses entanglement due to accumulated local noise, while stronger coupling $g$ enhances the entanglement harvesting. A notable feature is the non-monotonic dependence on $\omega \sigma$ : negativity initially increases due to better spectral overlap with field modes, but eventually decreases as high-frequency modes contribute weakly to long-range correlations due to their shorter wavelengths. This behaviour parallels the threshold phenomenon discussed in \cite{Zych:2010yk}, where entanglement between spatially separated regions vanishes beyond a critical scale, emphasising the spectral and spatial limitations in accessing vacuum entanglement.
%
\subsection{The  massless minimally coupled scalar field}
\begin{figure}
    \centering
    \includegraphics[width=1.0\linewidth]{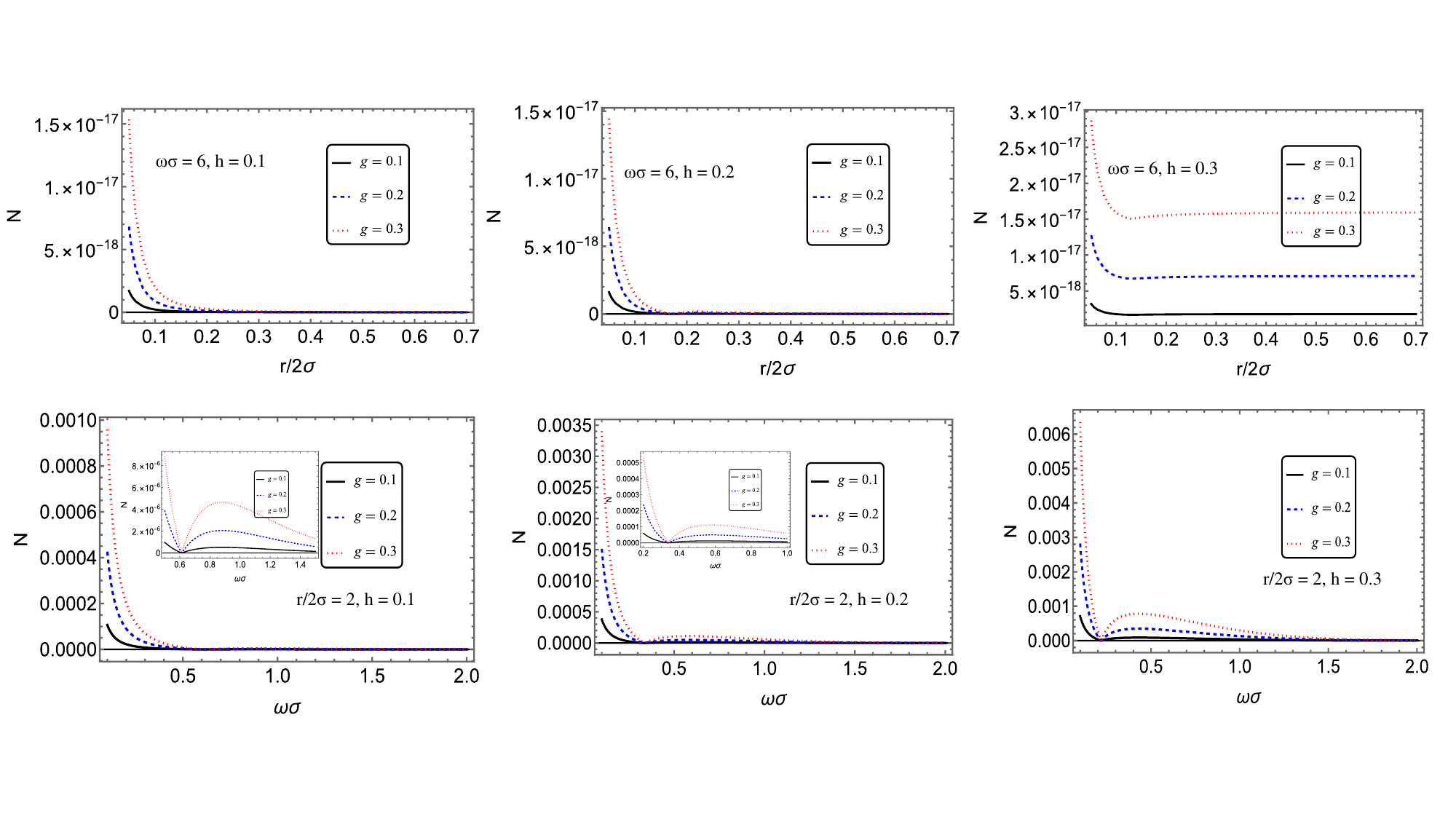}\vspace{-1.5cm}
    \caption{\small \it Negativity between two Unruh-DeWitt detectors coupled to a massless minimally coupled complex scalar field as a function of the dimensionless parameters $r/2\sigma$ and $\omega \sigma$, for different values of the coupling $g$ and interaction duration $h = H\sigma$, while keeping the infrared regulator fixed at $\Tilde{\epsilon}/H = e^{-20}$.
    }
    \label{NM}
\end{figure}

\noindent
Let us now come to the case of a massless, minimal scalar field, for which the Wightman function is given by \ref{WMCS}. Substituting \ref{WMCS} into \ref{PA} and \ref{E}, we may compute \( P_{IJ} \) and \( E \), similar to the earlier case of the conformal scalar. Let us first evaluate \( P_{IJ} \) for \( I = J \), where the spatial separation between the detectors is zero, i.e., \( r = 0 \), for which  \ref{WMCS} becomes
\begin{equation}
    \label{WMCSrzero}
    \begin{split}
 i G_M^+(x,x')&=\frac{H^2}{16\pi^2}[-\sinh^2{H(y-i\epsilon)}]^{-1}-\frac{H^2}{4\pi^2}{\rm Ei}\Big[-\frac{2i\Tilde{\epsilon}}{H}e^{-H\tau_+}\sinh{(H(y-i\epsilon)})\Big]+\frac{H^2}{4\pi^2}
   \end{split}
\end{equation}
We have
\begin{equation}
    \label{PMMSF}
\begin{split}
P&=\frac{H^2 g^2 h^2}{16\pi^4}\int_{-\infty}^{\infty} dx\; e^{-x^2} \int_{-\infty}^{\infty} dy\; e^{-y^2-2i\omega \sigma y}  \Bigg[\frac{1}{4\sinh^2{(h(y-i\epsilon))}}+{\rm Ei}\Big[-\frac{2i\Tilde{\epsilon}}{H}e^{-hx}\sinh{(h(y-i\epsilon))}\Big]-1\Bigg]^2
\end{split}\end{equation}
Upon expanding the square bracket and simplifying a bit, we have (see \ref{A1} for detail)
\begin{equation}
    \label{PMMSF1}
    \begin{split}
        P&=\frac{H^2 g^2 h^2}{8\pi^3}\int_{0}^{\infty}dk\;e^{-k^2}k\frac{\cosh{\frac{k\pi}{h}(1-2h\omega \sigma/\pi)}}{\sinh{\frac{k\pi}{h}}}+\frac{H^2 g^2 h^2 }{16\pi^3}e^{-\omega^2 \sigma^2}\\&+\frac{H^2 g^2 h^2 e^{-\omega^2 \sigma^2}}{96 \pi^3} \int_{0}^{\infty}dk\;e^{-k^2}k^3\frac{\cosh{\frac{k\pi}{h}(1-2h\omega \sigma/\pi)}}{\sinh{\frac{k\pi}{h}}}\\&+\frac{g^2H^2h^2 e^{-\omega^2\sigma^2}}{16\pi^3}\Bigg[\Bigg(\gamma+\ln\frac{2\Tilde{\epsilon}}{H}+\ln{(h(\omega + \epsilon))}+\frac{h^2}{2}\Bigg)^2 + \frac{h^2}{2}\Bigg]\\&+\frac{g^2H^2h^2 e^{-\omega^2\sigma^2}}{16\pi^3}\Bigg(\gamma+\ln\frac{2\Tilde{\epsilon}}{H}+\ln{(h(\omega + \epsilon))}+\frac{h^2}{2}\Bigg)+\frac{g^2H^2 e^{-\omega^2\sigma^2}}{64\pi^3 \Omega^2}\Bigg(\gamma+\ln\frac{2\Tilde{\epsilon}}{H}+\ln{(h(\omega + \epsilon))}\Bigg)
    \end{split}
\end{equation}
Likewise, we find matrix element $E$ using \ref{E} and \ref{WMCS} as
\begin{equation}
\begin{split}
    \label{EMMSF}
    &E= -g^2 \int_{-\infty}^{\infty} dx \; e^{-x^2+2i\omega \sigma x} \int_{-\infty}^{\infty} dy \; e^{-y^2}\Bigg[\frac{h^2}{16\pi^2 (-\sinh^2{(h(y-i\epsilon))}+e^{2hx}(hr/2\sigma)^2)}+\frac{h^2}{4\pi^2}\\&-\frac{h^2}{8\pi^2}{\rm Ei}\Big[-\frac{i\Tilde{\epsilon}}{H}\Big(-Hr+2e^{-hx}\sinh{(h(y-i\epsilon))}\Big)\Big]-\frac{h^2}{8\pi^2}{\rm Ei}\Big[-\frac{i\Tilde{\epsilon}}{H}\Big(Hr+2e^{-hx}\sinh{(h(y-i\epsilon))}\Big)\Big]\Bigg]^2\\
    &=-\frac{H^4 g^2 h^2}{16 \pi^4} \int_{-\infty}^{\infty} dx \; e^{-x^2+2i\omega \sigma x} \int_{-\infty}^{\infty} dy \; e^{-y^2} \Bigg[\frac{1}{4(-\sinh^2{(h(y-i\epsilon))}+e^{2hx}(\frac{hr}{2\sigma})^2)}+1\\&-\frac{1}{2}{\rm Ei}\Big[-\frac{i\Tilde{\epsilon}}{H}\Big(-Hr+2e^{-hx}\sinh{(h(y-i\epsilon))}\Big)\Big]-\frac{1}{2}{\rm Ei}\Big[-\frac{i\Tilde{\epsilon}}{H}\Big(Hr+2e^{-hx}\sinh{(h(y-i\epsilon))}\Big)\Big]\Bigg]^2,
    \end{split}
\end{equation}
which after simplifying a bit, reads (see \ref{A1} for details)
\begin{equation}
    \label{EMapp1f}
    \begin{split}
    E &= -\frac{g^2H^2h^2 e^{-\omega^2\sigma^2}}{16\pi^{7/2}}\int_{-\infty}^{\infty}dy\;e^{-y^2}\Bigg[\frac{1}{4\big(h^2(y-i\epsilon)^2+e^{i2h\omega \sigma}(hr/2\sigma)^2\big)}+1\\&+\gamma+\frac{1}{2}\Big(\ln{\Big[\frac{i\Tilde{\epsilon}}{H}(-Hr+2e^{-ih\omega\sigma}h(y-i\epsilon))\Big]}+\ln{\Big[\frac{i\Tilde{\epsilon}}{H}(Hr+2e^{-ih\omega\sigma}h(y-i\epsilon))\Big]}\Big)\Bigg]
    \end{split}
\end{equation}
We could not simplify \ref{PMMSF1} and \ref{EMapp1f} any further analytically. Plugging these expressions next into \ref{neg}, we have numerically computed the entanglement negativity as earlier and have investigated its variation in~\ref{NM}, keeping the infrared regulator fixed at $\Tilde{\epsilon}/H = e^{-20}$, following \cite{Nambu:2013rta}. Just like the previous case of the conformal scalar, we note that the negativity remains nonzero even when the detectors are separated beyond their causal horizon ($r/2\sigma \ensuremath{>} 1$), demonstrating the presence of long-range correlations in the minimally coupled vacuum. The negativity decreases with increasing interaction duration ($h = H\sigma$) and detector energy gap ($\omega \sigma$). Thus, as earlier, we see that longer interactions suppress entanglement due to local noise, while larger energy gaps reduce the detectors' sensitivity to low-frequency  (and hence long wavelength) modes that mediate entanglement. We also note qualitative and quantitative differences in \ref{NM} from the conformally coupled scalar case shown in \ref{N1}. In particular, for the minimally coupled scalar field, while the harvested entanglement may decay more rapidly with increasing separation in some regimes, the negativity remains strictly positive throughout the parameter space examined and, in fact, can surpass that of the conformal case at larger separations and longer interaction durations (compare the second row of \ref{N1} and \ref{NM}). This highlights the enhanced robustness of entanglement harvesting in the minimal case, likely due to the infrared amplification of long-range correlations characteristic of the minimally coupled field.

An important distinction emerges when comparing our results for the minimally coupled complex scalar with previous studies involving \emph{real} scalars and linear couplings. Notably, \cite{Nambu:2013rta} demonstrated that for a real scalar field in de Sitter spacetime, entanglement becomes undetectable once the detector separation exceeds the Hubble horizon (i.e., $r / 2\sigma \gtrsim 1$), consistent with the onset of classical behaviour. In contrast, our analysis shows that for a complex scalar field with Hermitian quadratic coupling ($\phi^\dagger \phi$), the negativity remains nonzero even in the \emph{infrared regime} ($\omega \sigma \ll 1$), and persists across super-horizon separations ($r / 2\sigma \gtrsim 1$). This robustness arises from both the infrared enhancement characteristic of the minimally coupled field and the amplification of non-local correlations induced by the nonlinear interaction. These results suggest that entanglement harvesting is more resilient in scenarios involving complex fields and nonlinear couplings, underscoring the sensitivity of observable quantum correlations to the nature of the field and its interaction with detectors.
%
\section{Conclusion}
\label{Conclusion}
\noindent
In this work, we analysed entanglement harvesting by two identical, comoving Unruh-DeWitt detectors in de Sitter spacetime, interacting with each other via either a conformally invariant or a minimally coupled massless complex scalar field. We computed the negativity as a measure of entanglement and studied its dependence on detector separation, energy gap, and interaction duration.

Our results show that for the conformally coupled scalar field, entanglement can be harvested even at super-horizon separations, highlighting the persistence of non-local correlations in the de Sitter vacuum. However, the degree of entanglement is sensitive to both the energy gap and the interaction duration. In contrast, the massless minimally coupled complex scalar field exhibits a qualitatively different behaviour: the negativity remains strictly positive throughout the parameter space considered, including at super-horizon separations. This indicates that some entanglement is always harvested, although its magnitude also decreases with increasing energy gap and interaction time. Notably, at larger separations and longer interaction durations, the harvested entanglement can surpass that of the conformal case. This enhancement can be attributed to the infrared amplification of long-wavelength field correlations in the minimally coupled field. Importantly, in both cases, the entanglement decays smoothly with increasing separation or energy gap, without exhibiting a sharp cutoff at the horizon scale—unlike the abrupt suppression observed in earlier studies involving real scalar fields with linear coupling (cf. discussion towards the end of the preceding section).

These findings underline that entanglement harvesting is shaped not only by causal structure, but also by the quantum nature of the field and the choice of coupling. They also suggest that features such as the persistence or suppression of negativity carry imprints of the underlying field theory, offering a window into the entanglement structure of quantum fields in curved spacetime.

Future investigations could extend this analysis to other types of fields and couplings, explore non-inertial detector trajectories, or consider decoherence effects. In particular, studying the decoherence dynamics between detectors—whether initially entangled or unentangled—may provide further insight into the transition from quantum to classical behaviour in cosmological environments. Extensions incorporating critical slowing down \cite{Kaplanek:2019vzj, Burgess:2024eng} in multi-detector settings also present compelling directions for future research.

\appendix
\labelformat{section}{Appendix #1} 
\section{A brief sketch of the computation of \ref{PCfinal} and \ref{ECfinal}}\label{A}

\begin{figure}
    \centering
    \includegraphics[width=.8\linewidth]{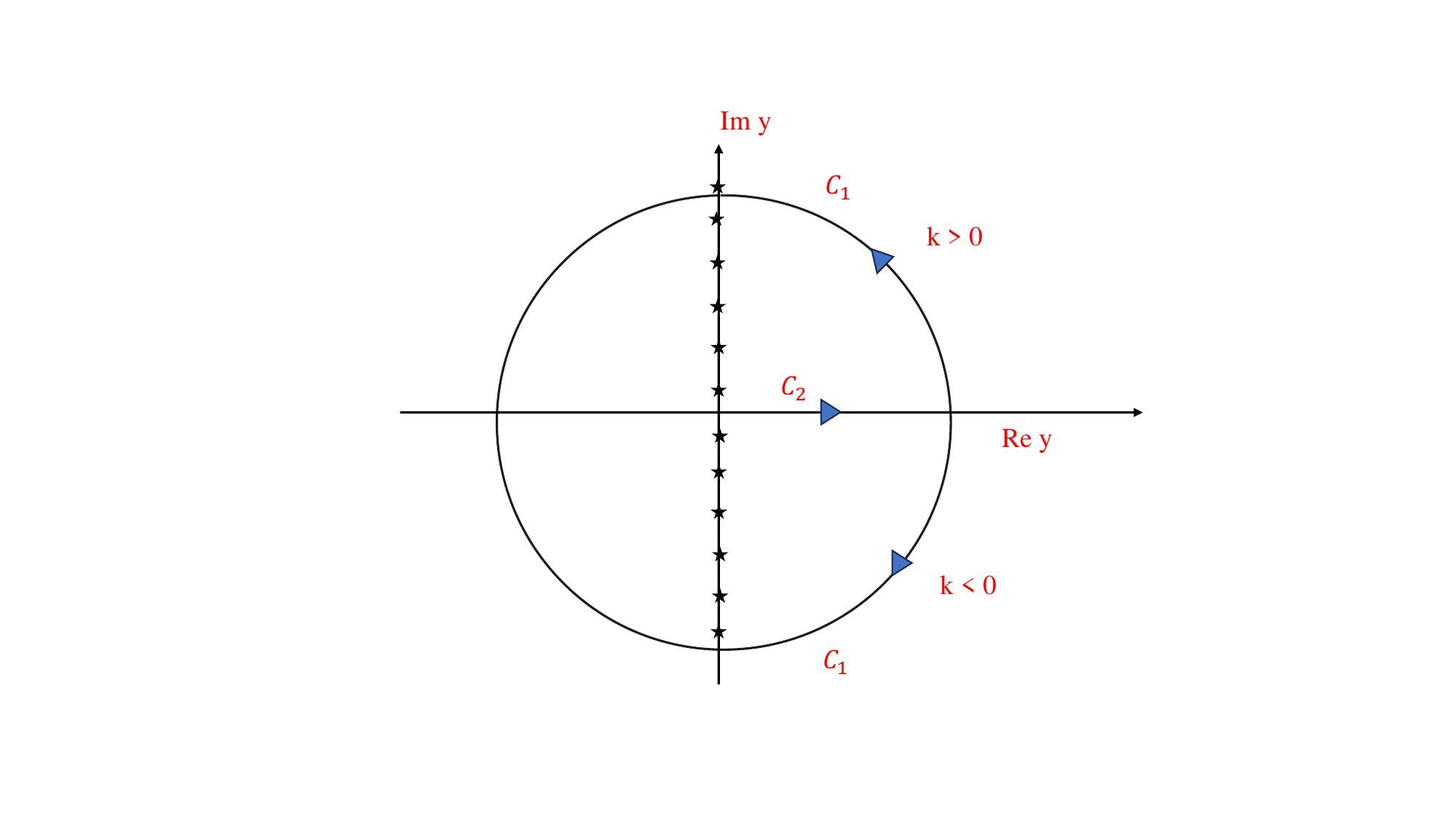}
    \caption{\small \it The integration contour for $A_1(k)$, \ref{A1'}. The radius of the contour $C_1$ is taken to be infinite.}
    \label{Contour1}
\end{figure}

\noindent
Following \cite{Nambu:2013rta}, we wish to sketch below some detail for the simplification of \ref{Pxy2} and \ref{Exy}, leading to \ref{PCfinal}, \ref{ECfinal}. For example, using
\begin{eqnarray}
e^{-y^2}=\frac{1}{\sqrt{\pi}}\int_{-\infty}^{\infty} dk\; e^{-k^2+2iky}
\label{n1}
\end{eqnarray}
in \ref{Pxy2}, we obtain
\begin{equation}
\begin{split}
    \label{n2}
    P&=\frac{H^2g^2 h^2}{256 \pi^4} \int_{-\infty}^{\infty}dk\; e^{-(k+\omega \sigma)^2}\int_{-\infty}^{\infty}dy\frac{e^{2iky}}{[\sinh{\big(h(y-i\epsilon)\big)}]^4}\\
    &=\frac{H^2g^2 h^2}{256 \pi^4}\int_{-\infty}^{\infty}dk\; e^{-(k+\omega \sigma)^2}\int_{-\infty}^{\infty}dy\sum_{n=-\infty}^{\infty} \frac{e^{2iky}}{\left(y-i\epsilon-\frac{i\pi n}{h}\right)^4}
    \end{split}
\end{equation}
it has $4^{th}$ order poles $4$ at $(i\pi n/h +i \epsilon)$. 
Let us consider the integral
\begin{equation}
    \label{A1'}
    A_1(k)= \int_{-\infty}^{\infty} dy \sum_{n=-\infty}^{\infty} \frac{e^{2iky}}{(y-i\epsilon-\frac{i\pi n}{h})^4}
\end{equation}
This integral can be solved by choosing the contour \ref{Contour1}. The residue at $i\epsilon+i\pi n$ is
\begin{equation}
    \label{Res}
\frac{1}{3!}\text{lim}_{\epsilon \to 0} \Big(\text{lim}_{y\to \frac{i \pi n}{h}+i\epsilon} \frac{d^3}{dy^3}(e^{2iky})\Big)=-\frac{8i}{6}e^{-\frac{2k\pi n}{h}}k^3
\end{equation}
On substituting \ref{Res} in \ref{A1'}, we obtain
\begin{equation}
 \label{A1first}
 A_1(k) 
 =\frac{8\pi k^3}{3(1-e^{-2\pi k/h})}
\end{equation}
Finally, on substituting \ref{A1first} into \ref{n2}, we get
\begin{equation}
\label{Psol}
  P =\frac{H^2 g^2 h^2}{96 \pi^3} \int_{-\infty}^{\infty} dk \;k^3\frac{ e^{-(k+\omega \sigma)^2}}{1-e^{-\frac{2\pi k}{h}}}
  = \frac{H^2 g^2 h^2e^{-\omega^2 \sigma^2}}{96\pi^3} \int_{0}^{\infty} dk \; e^{-k^2} k^3 \frac{\cosh{\big(\frac{k\pi}{h}(1-\frac{2h\omega \sigma}{\pi})\big)}}{\sinh{\frac{k\pi}{h}}}
\end{equation}
Next, in order to compute $E$, we substitute \ref{n1} into \ref{Exy} to obtain
\begin{eqnarray}
    \label{X1}
    E&=-\frac{H^2 g^2 h^2}{256 \pi^{5/2}}\int_{-\infty}^{\infty}\;dx e^{-x^2+2i\omega \sigma x} \int_{-\infty}^{\infty}dk\; e^{-k^2}  \int_{-\infty}^{\infty} dy\frac{e^{2iky}}{[-\sinh^2{\big(h(y-i \epsilon)\big)}+e^{2hx}(hr/2\sigma)^2]^2}\nonumber \\
    &=-\frac{H^2 g^2}{128 \pi^{5/2}}\int_{-\infty}^{\infty}dx\; e^{-x^2+2i\omega \sigma x} \int_{-\infty}^{\infty} dk\; e^{-k^2}  \int_{0}^{\infty} dy\frac{e^{2iky}}{\Big[\frac{\sinh^2{\big(h(y-i \epsilon)\big)}}{h^2}-e^{2hx}(r/2\sigma)^2\Big]^2}
\end{eqnarray}

\begin{figure}
    \centering
    \includegraphics[width=0.8\linewidth]{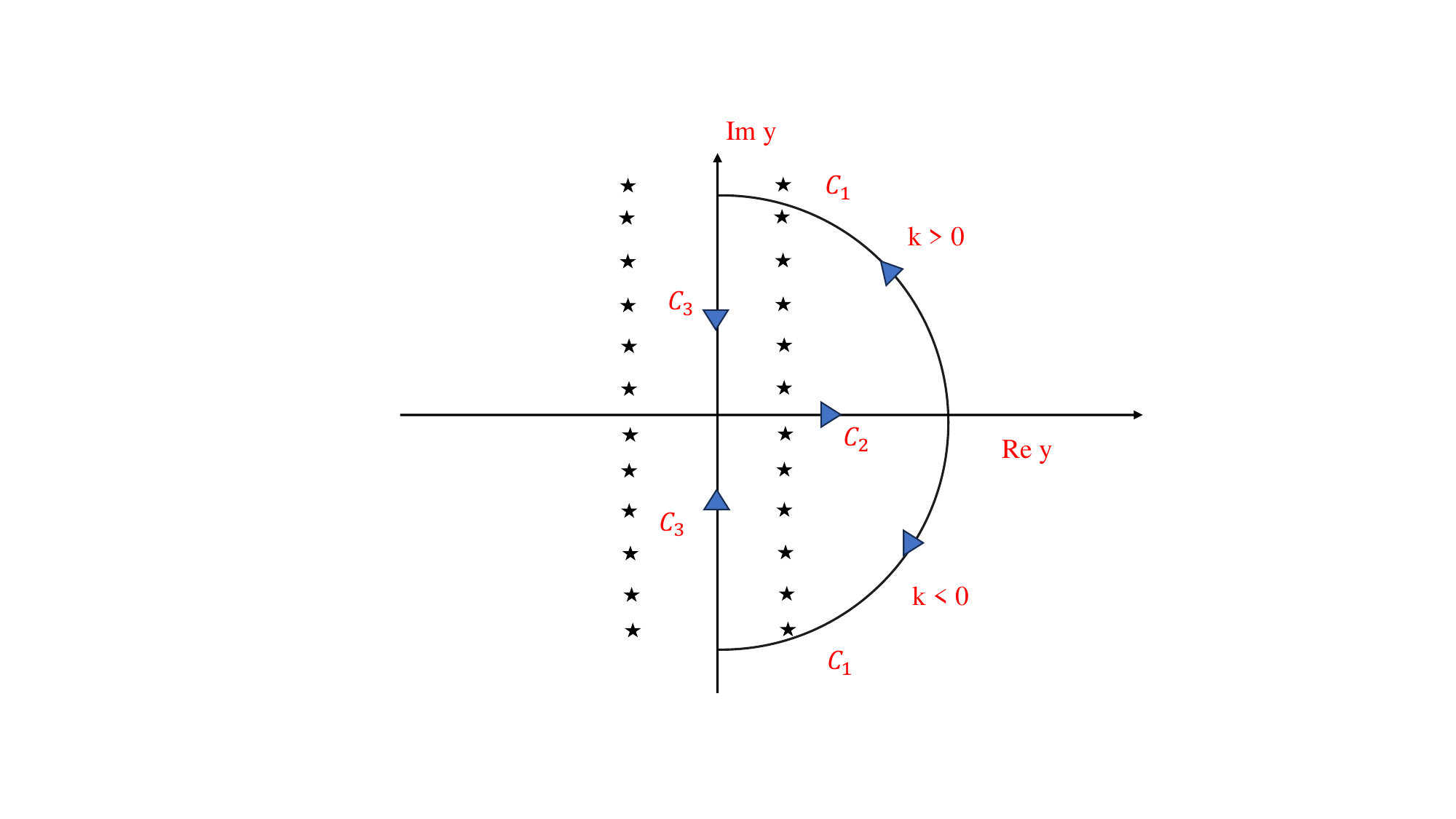}
    \vspace{-10mm}
    \caption{\small \it The integration contour for $A_2(k)$, \ref{A2b}.The radius of the contour $C_1$ is taken to be infinite.}
    \label{Contour2}
\end{figure}

\noindent
Let us now evaluate
\begin{equation}
    \label{A2}
    A_2(k)=\int_{0}^{\infty} dy \frac{e^{2iky}}{\Big[\frac{\sinh^2{\big(h(y-i \epsilon)\big)}}{h^2}-e^{2hx}(r/2\sigma)^2\Big]^2}
\end{equation}
Defining  $a=e^{hx}r/2\sigma$, we rewrite the above equation as
\begin{equation}
    \label{A2a}
    A_2(k)
    =\frac{1}{2a}\int_{0}^{\infty}dy\; e^{2iky} \Bigg(\frac{1}{\frac{\sinh{(h(y-i \epsilon))}}{h}-a}-\frac{1}{\frac{\sinh{(h(y-i \epsilon))}}{h}+a}\Bigg)^2
\end{equation}
The above integral can be rewritten as
\begin{equation}
\begin{split}
    \label{A2b}
    A_2(k)=\frac{1}{2a\big(ah+\frac{1}{b}\big)^2}\int_{0}^{\infty}dy\; e^{2iky} \sum_{n=-\infty}^{\infty}\Big[\frac{1}{(y-y_1)^2}+\frac{1}{(y-y_2)^2}-\frac{2}{(y-y_1)(y-y_2)}\Big]
    \end{split}
\end{equation}
where
\begin{equation}
\label{polesA2}
   y_1= \Big(i\epsilon+\frac{\ln{b}}{h}+\frac{in\pi}{h}\Big), \, \, y_2=\Big(i\epsilon-\frac{\ln{b}}{h}+\frac{in\pi}{h}\Big), \,\, b=ah+\sqrt{a^2h^2+1}.
\end{equation}
%
 %
%
%
Putting everything together, and choosing the contour depicted in~\ref{Contour2}, we obtain after a little algebra
\begin{equation}
    \label{A_2d}
    A_2(k)=\frac{2 \pi \sin{\Big(\frac{2k\ln{b}}{h}\Big)} \cosh{\frac{2\pi k}{h}}}{a(1+a^2h^2)(1-e^{\frac{2\pi k}{h}})}\Bigg(\frac{h}{\ln{b}}+2ik\Bigg)
\end{equation}
Finally, upon substituting \ref{A_2d} into \ref{X1}, we have
\begin{equation}
    \label{Ef}
    E=-\frac{H^2 g^2}{64 \pi^{3/2}}\int_{-\infty}^{\infty}dx\; e^{-x^2+2i\omega \sigma x} \int_{-\infty}^{\infty} dk\; e^{-k^2}  \frac{\sin{\Big(\frac{2k\ln{b}}{h}\Big)} \cosh{\frac{2\pi k}{h}}}{a(1+a^2h^2)\left(1-e^{\frac{2\pi k}{h}}\right)}\Bigg(\frac{h}{\ln{b}}+2ik\Bigg)
\end{equation}
%

\section{Asymptotic analysis of integrals \ref{Pxy2}, \ref{Exy}, \ref{PMMSF1} and \ref{EMMSF}}\label{A1}
\begin{figure}
    \centering
    \includegraphics[width=.9\linewidth]{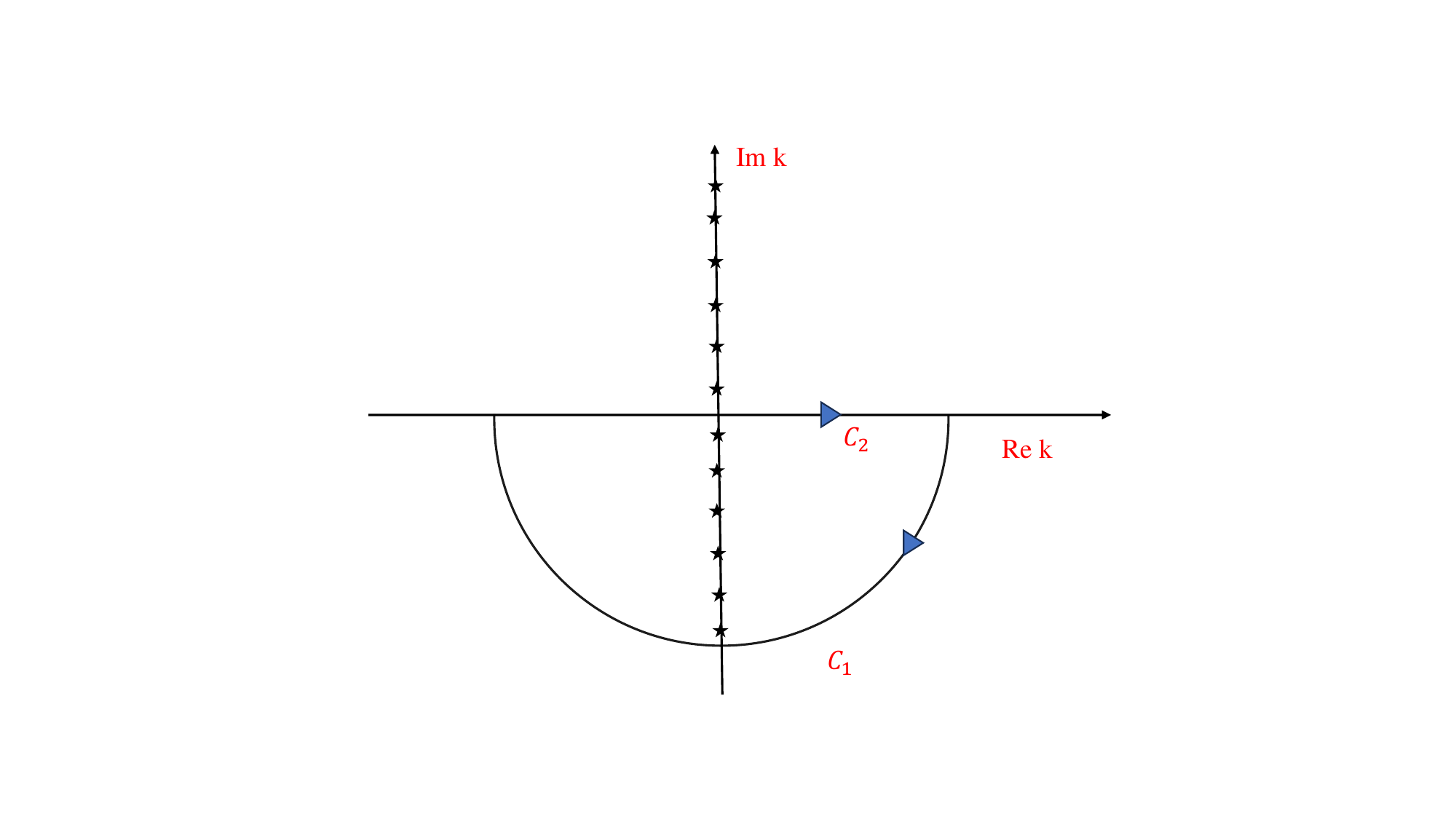}
    \vspace{-2.0cm}
    \caption{\it \small The integration contour for \ref{Ec}.}
    \label{contour3}
\end{figure}

\noindent
We introduce the following change of variables :
\begin{equation}
\label{approxredef}
    \omega \sigma \equiv \Omega,\quad \frac{h}{\pi}\equiv \Tilde{h}\quad \text{and} \quad \frac{r}{2\sigma} \equiv \Tilde{r}
\end{equation}
First, let us rewrite \ref{Pxy2} and \ref{Exy} respectively as,
\begin{equation}
    \label{Pxydef}
    P=\frac{H^2 g^2 \Tilde{h}^2 e^{-\Omega^2}}{96\pi}\int_{0}^{\infty} dk\;e^{-k^2}k^3\frac{\cosh{\frac{k(1-2\Tilde{h}\Omega)}{\Tilde{h}}}}{\sinh{\frac{k}{\Tilde{h}}}},
\end{equation}
and,
\begin{equation}
    \label{Exydef}
    E=-\frac{H^2g^2}{64\pi^{3/2}}\int_{-\infty}^{\infty}dx\;e^{-x^2+2i\Omega x}\int_{-\infty}^{\infty} dk\;e^{-k^2}\frac{\sin{\Big(\frac{2k}{\pi \Tilde{h}}\ln{b}\Big)}\cosh{\frac{2k}{\Tilde{h}}}}{a(1+a^2\Tilde{h}^2\pi^2)(1-e^{\frac{2k}{h}})}\Bigg(\frac{\Tilde{h}\pi}{\ln{b}}+2ik\Bigg),
\end{equation}
where $a$ and $b$  are defined as
\begin{equation}
\label{coeffdef}
    a=e^{\Tilde{h}\pi x}\Tilde{r}\quad \text{and} \quad b=a\pi\Tilde{h}+\sqrt{a^2\Tilde{h}^2\pi^2 + 1 } \end{equation}
We evaluated the integral in \ref{Pxydef} numerically. Next, following~\cite{Nambu:2013rta}, we proceed to evaluate \ref{Exydef} by first rewriting the approximations in \ref{ranges} in terms of the new variables introduced in \ref{approxredef}:

\begin{equation}
\label{approxdef}
    \Omega \gg 1, \qquad \frac{1}{\Tilde{h}}\gg 1 \;(\text{or}\; \Tilde{h} \ll 1)  \qquad \text{and} \qquad \Tilde{r} \gg  1
\end{equation}
Using these approximations, we evaluate the integral in equation~\ref{Exydef} via the saddle point method. Consider the integral
\begin{equation}
    \label{I}
    I = \int_{-\infty}^{\infty} dx\, e^{-x^2 + 2i\Omega x} F(x), \qquad \text{with} \quad \Omega \gg 1,
\end{equation}
The maximum of the phase of the exponential, $-x^2 + 2i\Omega x$ is located at $x = i\Omega$, i.e. on the imaginary axis. Expanding the phase around this maximum, we have  
\begin{equation}
    \label{I(iOmega)}
    I \approx e^{-\Omega^2} F(i\Omega) \int_{-\infty}^{\infty} dx\, e^{-(x - i\Omega)^2}
    = \sqrt{\pi}\, e^{-\Omega^2} F(i\Omega).
\end{equation}
We now rewrite~\ref{Exydef} as
\begin{equation}
    \label{Exysaddlept}
    E = -\frac{H^2 g^2e^{-\Omega^2}}{64 \pi}  \int_{-\infty}^{\infty} dk\; 
    \frac{
        e^{-k^2} 
        \sin\left( \frac{2k}{\pi \tilde{h}} \ln b(i\Omega) \right)
        \cosh\frac{2k}{h} 
    }{
        a(i\Omega) 
        \left( 1 + \pi^2 \tilde{h}^2 \ln a(i\Omega) \right) 
        \left( 1 - e^{2k/\tilde{h}} \right)
    }
    \left( \frac{\pi \tilde{h}}{\ln b(i\Omega)} + 2ik \right),
\end{equation}
where \( a(i\Omega) \) and \( b(i\Omega) \) denote the values of the functions \( a \) and \( b \) evaluated  at the saddle point \( x = i\Omega \).  We further rewrite \ref{Exysaddlept} as
\begin{equation}
E = -\frac{H^2 g^2}{64 \pi} e^{-\Omega^2}\int_{-\infty}^\infty  dk \frac{e^{-k^2} \sin\left( \frac{2k}{\pi \Tilde{h}} \log\left( e^{i \Tilde{h} \pi \Omega} \Tilde{r} \Tilde{h} + \sqrt{1 + (e^{i \Tilde{h} \pi \Omega} \Tilde{r})^2 \pi^2} \right) \right) \cosh\frac{2k}{\Tilde{h}} }{e^{i \Tilde{h} \pi \Omega} \Tilde{r} \left( 1 + (e^{i \Tilde{h} \pi \Omega} \Tilde{r})^2 \pi^2 \right) \left(1 - e^{2k/\Tilde{h}}\right)} 
\label{Ec}
\end{equation}
We integrate the above equation using the contour shown in \ref{contour3}. The integral is evaluated by summing the residues at the poles \( k_n = i \pi n \Tilde{h} \) for \( n < 0 \), which arise from the zeros of the denominator \( 1 - e^{2k/\Tilde{h}} = 0 \). The residue at each pole simplifies to
\begin{equation}
\mathrm{Res}_{k = k_n} f(k) \approx \frac{i \Tilde{h}}{2} \frac{\sinh\left( 2 n \log\left( e^{i \Tilde{h} \pi \Omega} \Tilde{r} \Tilde{h} + \sqrt{1 + (e^{i \Tilde{h} \pi \Omega} \Tilde{r})^2 \pi^2} \right) \right)}{e^{i \Tilde{h} \pi \Omega} \Tilde{r} \left( 1 + (e^{i \Tilde{h} \pi \Omega} \Tilde{r})^2 \pi^2 \right)}
\end{equation}
This yields
\begin{equation}
\label{Eapprox}
E \approx -\frac{H^2 g^2 \Tilde{h}}{128}  e^{-\Omega^2}\frac{\coth\left( \log\left( e^{i \Tilde{h} \pi \Omega} \Tilde{r} \Tilde{h} + \sqrt{1 + (e^{i \Tilde{h} \pi \Omega} \Tilde{r})^2 \pi^2} \right) \right)}{e^{i \Tilde{h} \pi \Omega} \Tilde{r} \left( 1 + (e^{i \Tilde{h} \pi \Omega} \Tilde{r})^2 \pi^2 \right)} 
\end{equation}
Furthermore, we numerically computed the  value of \ref{Eapprox} and the integral in \ref{Pxydef}, and subsequently obtained the negativity for the conformal scalar field using \ref{neg}. See main text for the same.\\

\noindent
We next proceed to analyse the integrals \ref{PMMSF1} and \ref{EMMSF} using the approximation specified in \ref{ranges} (\ref{approxredef}). We begin with \ref{PMMSF1}, which can be  expanded  as
\begin{equation}
\label{PMMexpn}
\begin{split}
 P&=\frac{g^2 H^2 h^2}{256\pi^4} \int_{-\infty}^{\infty}dx\; e^{-x^2}\int_{-\infty}^{\infty}dy \frac{e^{-y^2-2i\omega \sigma y}}{\big[\sinh({h(y-i\epsilon)})\big]^4} \\&+ \frac{g^2 H^2 h^2}{16\pi^4}\int_{-\infty}^{\infty}dx\; e^{-x^2}\int_{-\infty}^{\infty}dy\; e^{-y^2-2i\omega \sigma y}-\frac{g^2H^2h^2}{64\pi^4}\int_{-\infty}^{\infty}dx\; e^{-x^2}\int_{-\infty}^{\infty}dy \frac{e^{-y^2-2i\omega \sigma y}}{\sinh({h(y-i\epsilon)})}\\&+\frac{g^2 H^2 h^2}{16 \pi^4}\int_{-\infty}^{\infty}dx\;\int_{-\infty}^{\infty}dy\;e^{-x^2-y^2-2i\omega\sigma y} \Big[{\rm Ei}\Big(-\frac{2i\Tilde{\epsilon}}{H}e^{-hx}\sinh{(h(y-i\epsilon)}\Big)\Big]^2\\&-\frac{g^2 H^2 h^2}{16 \pi^4}\int_{-\infty}^{\infty}dx\;\int_{-\infty}^{\infty}dy\;e^{-x^2-y^2-2i\omega\sigma y} {\rm Ei}\Big(-\frac{2i\Tilde{\epsilon}}{H}e^{-hx}\sinh{(h(y-i\epsilon)}\Big)\\&+\frac{g^2 H^2 h^2}{64 \pi^4}\int_{-\infty}^{\infty}dx\;\int_{-\infty}^{\infty}dy\;e^{-x^2-y^2-2i\omega\sigma y} \frac{{\rm Ei}\Big(-\frac{2i\Tilde{\epsilon}}{H}e^{-hx}\sinh{(h(y-i\epsilon)}\Big)}{\big[\sinh({h(y-i\epsilon)})\big]^2}
\end{split}
\end{equation}
We solve the above expression term by term. The first term in \ref{PMMexpn} is identical to \ref{Pxy2}, which has been solved earlier and is given by \ref{Psol}. The second term can be evaluated using  Gaussian integral, 
\begin{equation}
    \label{term2}
    \frac{g^2 H^2 h^2}{16\pi^4} \int_{-\infty}^{\infty} dx\, e^{-x^2} \int_{-\infty}^{\infty} dy\, e^{-y^2 - 2i \omega \sigma y} = \frac{g^2 H^2 h^2}{16\pi^3} e^{-\omega^2 \sigma^2}
\end{equation}
The third term can be evaluated in a manner similar to \ref{Pxy2}, as shown in \ref{A}. We obtain
\begin{equation}
    \label{term3}
    -\frac{g^2 H^2 h^2}{64\pi^4} \int_{-\infty}^{\infty} dx\, e^{-x^2} \int_{-\infty}^{\infty} dy\, \frac{e^{-y^2 - 2i \omega \sigma y}}{\sinh(h(y - i\epsilon))} = \frac{g^2 H^2}{8\pi^3} \int_0^{\infty} dk\, e^{-k^2} k\, \frac{\cosh \left( \frac{\pi k}{h} \left(1 - \frac{2h\omega\sigma}{\pi} \right) \right)}{\sinh \frac{\pi k}{h} }
\end{equation}
Furthermore, to solve the remaining terms, we use the following approximation for the exponential integral function~\cite{GR} :
\begin{equation}
    \label{Eiapprox}
    {\rm Ei}[-z] \approx -\gamma - \ln z, \qquad \text{for} \; z \ll 1,
\end{equation}
Using the approximation \ref{Eiapprox} along with the saddle-point method discussed earlier in this Appendix to evaluate the integral \( I \), \ref{I(iOmega)}, we can compute the remaining terms, which are given by
\begin{equation}
\begin{split}
   \label{4thterm}
    \frac{g^2 H^2 h^2}{16 \pi^4}\int_{-\infty}^{\infty}dx\;\int_{-\infty}^{\infty}dy\;&e^{-x^2-y^2-2i\omega\sigma y}\Big[{\rm Ei}\Big(-\frac{2i\Tilde{\epsilon}}{H}e^{-hx}\sinh{(h(y-i\epsilon)}\Big)\Big]^2\\&=\frac{g^2H^2h^2 e^{-\omega^2\sigma^2}}{16\pi^3}\Bigg[\Bigg(\gamma+\ln\frac{2\Tilde{\epsilon}}{H}+\ln{(h(\omega + \epsilon))}+\frac{h^2}{2}\Bigg)^2 + \frac{h^2}{2}\Bigg]
    \end{split}
\end{equation}
\begin{equation}
\label{5thterm}
    \begin{split}
    -\frac{g^2 H^2 h^2}{16 \pi^4}\int_{-\infty}^{\infty}dx\;\int_{-\infty}^{\infty}dy\;&e^{-x^2-y^2-2i\omega\sigma y}{\rm Ei}\Big(-\frac{2i\Tilde{\epsilon}}{H}e^{-hx}\sinh{(h(y-i\epsilon)}\Big)\\&=\frac{g^2H^2h^2 e^{-\omega^2\sigma^2}}{16\pi^3}\Bigg(\gamma+\ln\frac{2\Tilde{\epsilon}}{H}+\ln{(h(\omega + \epsilon))}+\frac{h^2}{2}\Bigg)
    \end{split}
\end{equation}
\begin{equation}
\label{6thterm}
    \begin{split}
    \frac{g^2 H^2 h^2}{64 \pi^4}\int_{-\infty}^{\infty}dx\;\int_{-\infty}^{\infty}dy\;&e^{-x^2-y^2-2i\omega\sigma y} \frac{{\rm Ei}\Big(-\frac{2i\Tilde{\epsilon}}{H}e^{-hx}\sinh{(h(y-i\epsilon)}\Big)}{\big[\sinh({h(y-i\epsilon)})\big]^2}\\&=\frac{g^2H^2 e^{-\omega^2\sigma^2}}{64\pi^3 \Omega^2}\Bigg(\gamma+\ln\frac{2\Tilde{\epsilon}}{H}+\ln{(h(\omega + \epsilon))}\Bigg)
    \end{split}
\end{equation}
Combining all these terms, we now obtain
\begin{equation}
    \label{PMMSF1A}
    \begin{split}
        P&=\frac{H^2 g^2 h^2}{8\pi^3}\int_{0}^{\infty}dk\;e^{-k^2}k\frac{\cosh{\big(\frac{k\pi}{h}(1-2h\omega \sigma/\pi)\big)}}{\sinh{\big(\frac{k\pi}{h}\big)}}+\frac{ g^2 H^2 h^2 }{16\pi^3}e^{-\omega^2 \sigma^2}\\&+\frac{ g^2 H^2 h^2 e^{-\omega^2 \sigma^2}}{96 \pi^3} \int_{0}^{\infty}dk\;e^{-k^2}k^3\frac{\cosh{\big(\frac{k\pi}{h}(1-2h\omega \sigma/\pi)\big)}}{\sinh{\big(\frac{k\pi}{h}\big)}}\\&+\frac{g^2H^2h^2 e^{-\omega^2\sigma^2}}{16\pi^3}\Bigg[\Bigg(\gamma+\ln\frac{2\Tilde{\epsilon}}{H}+\ln{(h(\omega + \epsilon))}+\frac{h^2}{2}\Bigg)^2 + \frac{h^2}{2}\Bigg]\\&+\frac{g^2H^2h^2 e^{-\omega^2\sigma^2}}{16\pi^3}\Bigg(\gamma+\ln\frac{2\Tilde{\epsilon}}{H}+\ln{(h(\omega + \epsilon))}+\frac{h^2}{2}\Bigg)+\frac{g^2H^2 e^{-\omega^2\sigma^2}}{64\pi^3 \Omega^2}\Bigg(\gamma+\ln\frac{2\Tilde{\epsilon}}{H}+\ln{(h(\omega + \epsilon))}\Bigg)
    \end{split}
\end{equation}
Afterwards, it can be solved numerically. Next, to compute \( E \)  given by \ref{EMMSF}, we proceed similarly to the calculation of \( P \). We use the saddle-point approximation and the approximation  given by \ref{Eiapprox}, to obtain
\begin{equation}
    \label{EMapp1}
    \begin{split}
    E&= -\frac{g^2H^2h^2 e^{-\omega^2\sigma^2}}{16\pi^{7/2}}\int_{-\infty}^{\infty}dy\;e^{-y^2}\Bigg[\frac{1}{4\big(h^2(y-i\epsilon)^2+e^{i2h\omega \sigma}(hr/2\sigma)^2\big)}+1\\&+\gamma+\frac{1}{2}\Big(\ln{\Big[\frac{i\Tilde{\epsilon}}{H}(-Hr+2e^{-ih\omega\sigma}h(y-i\epsilon))\Big]}+\ln{\Big[\frac{i\Tilde{\epsilon}}{H}(Hr+2e^{-ih\omega\sigma}h(y-i\epsilon))\Big]}\Big)\Bigg],
    \end{split}
\end{equation}
which we  have computed numerically.

\bibliographystyle{cas-model2-names}
\bibdata{outputNotes,outputNotes : ,cas-refs}

\end{document}